# (Sr,Na)(Zn,Mn)$_2$As$_2$: A new diluted ferromagnetic semiconductor with the hexagonal CaAl$_2$Si$_2$ type structure

B. J. Chen[1], K. Zhao[1], Z. Deng[1], W. Han[1], J. L. Zhu[1], X. C. Wang[1], Q. Q. Liu[1], B. Frandsen[2], L. Liu[2], S. Cheung[2], F. L. Ning[3], T.J.S. Munsie[4], T. Medina[4], G.M. Luke[4], J.P. Carl[5], J. Munevar[6], Y. J. Uemura[2] and C. Q. Jin[1]

1. Beijing National Laboratory for Condensed Matter Physics and Institute of Physics, Chinese Academy of Sciences, Collaborative Innovation Center of Quantum Matter, Beijing, China
2. Department of Physics, Columbia University, New York, New York 10027, USA
3. Department of Physics, Zhejiang University, Hangzhou 310027, China
4. Department of Physics & Astronomy, McMaster University, Hamilton, Canada.
5. Department of Physics, Villanova University, Villanova, PA 19085, USA
6. Centro Brasileiro de Pesquisas Fisicas, Rio de Janeiro, Brazil

**Abstract**

A new diluted ferromagnetic semiconductor (Sr,Na)(Zn,Mn)$_2$As$_2$ is reported, in which charge and spin doping are decoupled via Sr/Na and Zn/Mn substitutions, respectively, being distinguished from classic (Ga,Mn)As where charge & spin doping are simultaneously integrated. Different from the recently reported ferromagnetic (Ba,K)(Zn,Mn)$_2$As$_2$, this material crystallizes into the hexagonal CaAl$_2$Si$_2$-type structure. Ferromagnetism with a Curie temperature up to 20 K has been observed from magnetization. The muon spin relaxation measurements suggest that the exchange interaction between Mn moments of this new system could be different to the earlier DMS systems. This system provides an important means for studying ferromagnetism in diluted magnetic semiconductors.

## I. Introduction

Diluted magnetic semiconductors (DMS) take advantage of both the charge and spin of the electron, yielding remarkable properties and functionalities. The discovery of p-type Mn-doped DMS systems, especially (Ga,Mn)As, has led to the development of multifunctional materials that combine the capabilities of semiconductors and ferromagnets.[1,2,3] In the prototypical (Ga,Mn)As system, substitution of the magnetic ion $Mn^{2+}$ for $Ga^{3+}$ couples spin and charge doping simultaneously. This coupling precludes the possibility of individually tuning the spin and charge degrees of freedom. Additionally, the limited chemical solubility of Mn substituted for Ga restricts bulk specimens of (Ga,Mn)As to very low Mn concentrations (< 1%), although high-quality metastable thin films with Mn concentrations up to ~15% can be fabricated via molecular beam epitaxy[4,5]. These specimens tend to be highly sensitive to the particular preparation method and heat treatment. These challenges have led to significant interest in finding new types of DMS materials to overcome these obstacles.

Li(Zn,Mn)As was the first of a new class of DMS materials to be discovered that successfully decouples charge and spin doping[6]. This bulk material utilizes excess Li concentration to introduce mobile carriers, while independently exploiting the isovalent substitution of $Mn^{2+}$ for $Zn^{2+}$ to introduce local spins. Shortly after the discovery of Li(Zn,Mn)As, another new type bulk ferromagnetic DMS, $(Ba,K)(Zn,Mn)_2As_2$, was synthesized, crystallizing into the so-called "122" structure[7] and exhibiting a much higher Curie temperature ($T_C$) up to 180 K[8]. These discoveries initiated a fruitful exploration of DMS systems, with several new DMS materials discovered, including Li(Zn,Mn)P[9], (La,Ca)(Zn,Mn)SbO[10], (La,Ba)(Zn,Mn)AsO[11] and (La,Sr)(Cu,Mn)SO[12].

In this paper, we report the discovery of a new DMS compound, (Sr,Na)(Zn,Mn)$_2$As$_2$. This compound is a p-type semiconductor with the hexagonal CaAl$_2$Si$_2$ structure[13]. Ferromagnetism with a maximum $T_C$ ~20 K was achieved with decoupled charge and spin doping through Sr/Na and Zn/Mn substitutions, respectively.

**II. Experiment**

Polycrystalline specimens of (Sr$_{1-y}$Na$_y$)(Zn$_{1-x}$Mn$_x$)$_2$As$_2$ were synthesized using the solid state reaction method[14]. All the starting materials, SrAs and Na$_3$As precursors, high purity Zn and Mn powders, were mixed according to the nominal composition of (Sr$_{1-y}$Na$_y$)(Zn$_{1-x}$Mn$_x$)$_2$As$_2$. The mixtures were pressed into pellets and sealed inside evacuated tantalum tubes to prevent the evaporation of Na. Then the tantalum tubes were sealed inside evacuated quartz tubes. The tubes were heated to 1000 K and held for several days before the furnace slowly cooled to room temperature. The specimens were characterized by X-ray powder diffraction with a Philips X'pert diffractometer using Cu K$_\alpha$ radiation at room temperature. Lattice parameters were determined via Rietveld analysis using the GSAS software package[15]. DC magnetic susceptibility measurements were performed on a Quantum Design SQUID VSM at temperatures of 2 to 300 K, and electrical transport measurements were carried out on a Quantum Design PPMS in the same temperature range. Muon spin relaxation (μSR) measurements were performed at TRIUMF, in Vancouver, Canada.

**III Results and discussion**

As shown in Fig. 1(a), all of the main peaks in the X-ray diffraction patterns can be indexed by the hexagonal CaAl$_2$Si$_2$ type structure with the space group P-3m1, excluding a few minor peaks of nonmagnetic $Sr_3As_4$ and $Zn_3As_2$ impurities. Although the crystal structure of the present material is very different from those of the recently reported "111", "122" and "1111" DMS materials[6, 16~19], all of these systems have a layered structure and share the common feature of [$ZnAs_4$] tetrahedral coordination. Here the [$ZnAs_4$] tetrahedron is not regular, since the Zn-As bond along the c-axis ($L_c$) is longer than the other three Zn-As bonds ($L_{ab}$) and the As-Zn-As bond angle $\alpha$ is bigger than $\beta$, as shown in Fig. 1(b). As seen in Fig. 1(c), the lattice parameters of $(Sr_{0.9}Na_{0.1})(Zn_{1-x}Mn_x)_2As_2$ for x= 0, 0.05, 0.1, 0.15 and 0.2 increase monotonically with $x$, indicative of successful chemical doping. The parent compound $SrZn_2As_2$ has lattice parameters $a$ = 4.223 Å and $c$ = 7.268 Å. The isovalent doping of $Mn^{2+}$ for $Zn^{2+}$ allows the synthesis of chemically stable bulk crystals of $(Sr,Na)(Zn_{1-x}Mn_x)_2As_2$ for $x$ up to 0.2.

The temperature-dependent magnetization $M(T)$ for $(Sr_{1-y}Na_y)(Zn_{1-x}Mn_x)_2As_2$ specimens are displayed in Fig.2(a). Clear signatures of ferromagnetism are observed. The solid symbols in Fig. 2(a) show $M(T)$ for $(Sr_{0.9}Na_{0.1})(Zn_{1-x}Mn_x)_2As_2$ with $x$ = 0.05, 0.1, 0.15, and 0.2. The Curie temperature $T_C$ rises with increasing Mn concentration, reaching a maximum of 21 K. The empty symbols show $T_C$ of $(Sr_{1-y}Na_y)(Zn_{0.85}Mn_{0.15})_2As_2$ as function of Na concentration $y$. The maximum $T_C$ is 24K for optimal Na doping ($y$= 0.2). Further Na doping causes the Curie temperature to gradually decrease. Above $T_C$, the samples are paramagnetic and the magnetic susceptibility $\chi(T)$ (Fig. 2(a) and Fig.2(b)) can be fit with the Curie-Weiss formula $(\chi-\chi_0)^{-1} = (T-\theta)/C$, quite well in high temperature region (30K ~ 300 K) as shown in Fig. 2(a). In this formula, $\chi_0$ is a temperature-independent paramagnetic term that depends on the orbital contribution of the material[20], $C$ is the Curie constant, and $\theta$ is the Weiss temperature. The positive value of $\theta$ for $(Sr_{1-y}Na_y)(Zn_{0.8}Mn_{0.2})_2As_2$ (Fig. 2(a)) indicates a ferromagnetic interaction between $Mn^{2+}$ ions. The effective paramagnetic moment ($M_{eff}$) obtained from the Curie constant (open purple circles in Fig. 2(b)) decreases with increasing Mn doping. A similar trend was also found in other systems doped with magnetic ions[6,8,11]. $M_{eff}$ of the specimen $(Sr_{0.9}Na_{0.1})(Zn_{0.95}Mn_{0.05})_2As_2$ is close to $5.9\mu_B$ per Mn. According to $M_{eff} = g[S(S+1)]^{1/2}$, where g = 2, the value 5.9 $\mu_B$ corresponds to S = 5/2, confirming the divalent status of the Mn ion where five d-electrons are fully high-spin oriented. Figure 2(b) also shows the saturation moment ($M_{sat}$) per Mn in an applied field of 1000 Oe. $M_{sat}$ of all specimens is 1–2 $\mu_B$/Mn, comparable to that of (Ga,Mn)As, Li(Zn,Mn)As, (Ba,K)(Zn,Mn)$_2$As$_2$ and Li(Zn,Mn)P[1,6,8,9]. Fig. 2(b) also shows that $M_{sat}$ per Mn decreases with increasing Mn concentration, probably due to competition between antiferromagnetic nearest-neighbor interactions and ferromagnetic interactions of Mn moments since there are two type of exchange interactions with one favorable for ferromagnetic order for low Mn concentration while the other for antiferromagnetc order by superexchange interaction for neighbor Mn[6]. The hysteresis curves of $(Sr_{1-y}Na_y)(Zn_{0.85}Mn_{0.15})_2As_2$ at $T$ = 2 K are displayed in Fig.2(c). The coercive field ($H_C$) decreases from 110 Oe to 10 Oe with increasing Na doping. This small coercive field could be advantageous for spin flip manipulation.

To determine $T_C$ accurately, Arrott-Noakes plots[21,22] (that is, the square of the magnetization $M^2$ versus the ratio of the applied magnetic field and the magnetization, $H/M$) were produced for all samples. The isothermal $M$-$H$ measurements of $Sr_{0.8}Na_{0.2}(Zn_{0.85}Mn_{0.15})_2As_2$ seen in Fig.2(d) were re-drawn as plot of $M^2$ versus $H/M$ over the temperature range of 5-35K, shown in Fig.2(e). According to Weiss-Brillouin molecular field theory, around the Curie point[21],

$$H/M = \alpha(T-T_C) + \beta M^2 + \gamma M^4 + \ldots \quad (1)$$

It can be seen from Fig. 2(e) that isotherms above the calculated Curie point have a positive intercept, while the isotherms below the calculated Curie point have a negative intercept. The isotherm at the Curie point is a straight line passing through the origin. In this way, the Curie temperature was determined from Fig. 2(e) to be 20K for this composition. The Curie temperature for the other samples was obtained similarly.

Results of electronic transport studies are shown in Fig. 3. Resistivity measurements shown in Fig. 3(a) indicate that $SrZn_2As_2$ is a semiconductor. Substituting Na for Sr atoms introduces hole carriers, decreasing the resistivity and leading to metallic behavior in $(Sr,Na)Zn_2As_2$. As seen in Figure 3(a), the resistivity of $(Sr_{0.9}Na_{0.1})(Zn_{1-x}Mn_x)_2As_2$ increases monotonically with increasing Mn concentration, suggesting that Mn acts as a scattering centre[6]. The specimens show a local maximum of resistance around the Curie temperature, with decreasing resistivity below $T_c$. This critical behavior of $\rho$ is commonly observed in other magnetic semiconductors[6,23]. As the temperature is lowered from above $T_C$, the resistivity increases to the local maximum around $T_C$ as a consequence of critical scattering, in which carriers are scattered by correlated spin fluctuations resulting from short-range interactions between spins[22,23]. The resistance then decreases below $T_C$ because spin scattering is reduced in the ferromagnetic state due to the alignment of magnetic moments. The details of the evolution of the ferromagnetic order will be discussed later.

Fig. 3(b) shows the resistivity curves $\rho(T)$ of $(Sr_{0.8}Na_{0.2})(Zn_{0.85}Mn_{0.15})_2As_2$ at various magnetic fields. Similar to the case of (Ga,Mn)As[22], the local maximum of resistance around the Curie temperature moved to higher temperature with the increasing applied magnetic fields. Fig. 3(c) shows magnetotransport measurements of $(Sr_{0.8}Na_{0.2})(Zn_{0.85}Mn_{0.15})_2As_2$ at various temperatures. The negative magnetoresistance does not saturate even in a rather high field, where the spins are fully aligned according to M(*H*) (Fig. 2(c)). In this case, the negative magnetoresistance is presumably due more to the effects of weak localization than the reduction of spin-dependent scattering[24].

The anomalous Hall effect (AHE) resulting from spontaneous magnetization is strong evidence for ferromagnetism in this DMS system. AHE is generated in ferromagnetic ordered materials. Below the Curie temperature, the AHE is clearly observed in $(Sr_{0.8}Na_{0.2})(Zn_{0.85}Mn_{0.15})_2As_2$, as displayed in Fig. 3(d). The linear parts of the Hall resistivity were used to ascertain the presence of p-type carriers with a concentration of $n_p=1.09\times10^{20}$ cm$^{-3}$, about 10% of the nominal concentration of Na. Hole carriers are expected from the substitution of Na$^+$ for Sr$^{2+}$. Assuming each single Na$^+$ donates one hole, 20% Na doping should result in a hole concentration of approximately $1.8\times10^{21}$ cm$^{-3}$, so most part of them should have been compensated by defects. A similar result of the hole concentration has been obtained in (Ga,Mn)As system, in which most of the Mn centers which are acting as acceptors have been compensated by As antisite defects[25-27]. This carrier concentration is comparable to that of (Ga,Mn)As[1], Li(Zn,Mn)As[6,] (Ba,K)(Zn,Mn)$_2$As$_2$[8] and (La,Ca)(Zn,Mn)SbO[11].

μSR is an ideal probe to examine volume fraction and the ordered moment size in DMS systems[28-30]. Bulk polycrystalline specimens were used for μSR measurements, confirming the presence of ferromagnetism in (Sr,Na)(Zn,Mn)$_2$As$_2$. Fig. 4(a) shows the zero-field (ZF) μSR time spectra for (Sr$_{0.8}$Na$_{0.2}$)(Zn$_{0.85}$Mn$_{0.15}$)$_2$As$_2$ at different temperatures, with the relaxation rates plotted in the inset of Fig. 4(b). The onset of rapid relaxation below 26 K indicates the appearance of bulk magnetic order in this system with a Curie temperature between 22 K and 26 K, in agreement with the magnetization results. The temperature evolution of the relaxation rate, which is proportional to the size of the ordered moment, indicates a steady increase in moment size as the temperature is lowered. The total amplitude of the fast-relaxing component of the asymmetry is proportional to the volume fraction of the magnetically ordered phase. As seen in Fig. 4(b), the ordered volume fraction gradually increases below $T_C$ to 100%, matching well with the spontaneous magnetization below $T_C$. Fig. 4(c) compares the present result on Sr$_{0.8}$Na$_{0.2}$(Zn$_{0.85}$Mn$_{0.15}$)$_2$As$_2$ with the earlier DMS systems of (Ga,Mn)As[30] system, the 111-type system Li(Zn,Mn)[6], the 122-type system (Ba,K)(Zn,Mn)$_2$As$_2$[8] and the 1111-type system (La,Ba)(Zn,Mn)AsO[10] in a plot of the low-temperature relaxation rate $a_s$ versus the Curie temperature $T_C$. The data point representing the current work comes to a very different location as compared to all the earlier DMS systems, which tend to lie along a common line. This suggests that exchange interaction between Mn moments of this new 122-type DMS with hexagonal CaAl$_2$Si$_2$-type structure could be different than that of Ba122 and other DMS systems.

## IV. Conclusions

In summary, a new bulk ferromagnetic DMS material, $(Sr_{1-y}Na_y)(Zn_{1-x}Mn_x)_2As_2$, with a layered hexagonal structure was synthesized. Together with carrier doping via (Sr,Na) substitution, spin doping via (Zn,Mn) substitution results in ferromagnetic order with a Curie temperature up to 20 K. The coercive field of this p-type material is less than 110 Oe. This system will motivate further exploration of new DMS systems in hexagonal semiconductors.

**Acknowledgments**: The work was supported by NSF &MOST through Research Projects.

# References

[1] H. Ohno, **Science 281**, 951 (1998).

[2] T. Dietl, **Nat. Mater. 9**, 965 (2010).

[3] I. Zutic, J. Fabian, and S. Das Sarma, **Rev. Mod. Phys.76**, 323 (2004).

[4] H. Ohno, A. Shen, F. Matsukura, A. Oiwa, A. Endo, S. Katsumoto, and Y. Iye, **Appl. Phys. Lett. 69**, 363 (1996).

[5] T. Dietl, H. Ohno, F. Matsukura, J. Cibert, and D. Ferrand, **Science 287**, 1019 (2000).

[6] Z. Deng, C. Q. Jin, Q. Q. Liu, X. C. Wang, J. L. Zhu, S. M. Feng, L. C. Chen, R. C. Yu, C. Arguello, T. Goko, F. Ning, J. Zhang, Y. Wang, a a Aczel, T. Munsie, T. J. Williams, G. M. Luke, T. Kakeshita, S. Uchida, W. Higemoto, T. U. Ito, B. Gu, S. Maekawa, G. D. Morris, and Y. J. Uemura, **Nat. Commun. 2,** 422 (2011).

[7] M. Rotter, M. Tegel, I. Schellenberg, W. Hermes, R.Pottgen, and D. Johrendt, **Phys. Rev. B 78**, 20503(R) (2008).

[8] K. Zhao, Z. Deng, X. C. Wang, W. Han, J. L. Zhu, X. Li, Q. Q. Liu, R. C. Yu, T. Goko, B. Frandsen, L. Liu, F. Ning, Y. J. Uemura, H. Dabkowska, G. M. Luke, H. Luetkens, E. Morenzoni, S. R. Dunsiger, A. Senyshyn, P. Böni, and C. Q. Jin, **Nat. Commun. 4,** 1442 (2013).

[9] Z. Deng, et al., **Phys. Rev. B 88,** R081203 (2013).

[10] C. Ding, et al., **Phys. Rev. B 88**, R041102 (2013).

[11] W. Han, et al., **Science China Physics 56**, 2026 (2013).

[12] X.J. Yang, Y.K. Li, C.Y. Shen, B.Q. Si, Y.L. Sun, Q. Tao, G.H. Cao, Z.A. Xu, and F. C. Zhang, **Appl. Phys. Lett. 103**, 022410 (2013).

[13] C. Zheng, R. Hoffmann, R. Nesper, and H. G. Von Schnering, **J. Am. Chem. Soc. 108**, 1876 (1986).

[14] K. Zhao, Q. Q. Liu, X. C. Wang, Z. Deng, Y. X. Lv, J. L. Zhu, F. Y. Li, and C. Q. Jin, **Phys. Rev. B 84**, 184534 (2011)

[15] A. C. Larson and R. B. Von Dreele, General Structure Analysis System. LANSCE, Los Alamos, LAUR 86-748 (1994).

[16] Y. Kamihara, T. Watanabe, M. Hirano, and H. Hosono, **J. Am. Chem. Soc.130**, 3296(2008).

[17] M. Rotter, M. Tegel, and D. Johrendt, **Phys. Rev. Lett. 101**, 107006 (2008).

[18] X. Wang, Q. Liu, Y. Lv, W. Gao, L. Yang, R. Yu, F. Li, and C. Jin, **Solid State Commun. 148**, 538 (2008).

[19] Z. Deng, X. Wang, Q. Liu, S. Zhang, Y. Lv, J. Zhu, R. Yu, and C. Jin, **Europhysics Lett. 87**, 37004 (2009).

[20] A. S. Erickson, S. Misra, G. J. Miller, R. R. Gupta, Z. Schlesinger, W. A. Harrison, J. M. Kim, and I. R. Fisher, **Phys. Rev. Lett. 99,** 016404 (2007).

[21] A. Arrott, **Phys. Rev. 108**, 1394 (1957).

[23] V. Novak, et al., **Phys. Rev. Lett. 101** (2008).

[22] F. Matsukura, H. Ohno, A. Shen, and Y. Sugawara, **Phys. Rev. B 57**, R2037 (1998).

[24] H. Ohno, A. Shen, F. Matsukura, A. Oiwa, A. Endo, S. Katsumoto, and Y. Iye, **Appl. Phys. Lett. 69**, 363 (1996).

[25] M. Luysberg, H. Sohn, A. Prasad, P. Specht, Z. Liliental-Weber, E. R. Weber, J. Gebauer, and R. Krause-Rehberg, **J. Appl. Phys. 83**, 561 (1998).

[26] T. Omiya, F. Matsukura, T. Dietl, Y. Ohno, T. Sakon, M. Motokawa, and H. Ohno, **Physica E 7**, 976 (2000).

[27] T. Hayashi, Y. Hashimoto, S. Katsumoto, and Y. Iye, **Appl. Phys. Lett. 78**, 1691 (2001).

[28] Y. Uemura, et al., **Nat. Phys. 3**, 29 (2006).

[29] Y. Uemura, T. Yamazaki, D. Harshman, M. Senba, and E. Ansaldo, **Phys. Rev. B 31**, 546 (1985).

[30] S. R. Dunsiger, J. P. Carlo, T. Goko, G. Nieuwenhuys, T. Prokscha, A. Suter, E. Morenzoni, D. Chiba, Y. Nishitani, T. Tanikawa, F. Matsukura, H. Ohno, J. Ohe, S. Maekawa, and Y. J. Uemura, **Nat. Mater. 9**, 299 (2010).

Figure Captions:

**Fig.1**. (a) X-ray diffraction patterns for $(Sr_{0.9}Na_{0.1})(Zn_{1-x}Mn_x)_2As_2$ at room temperature. The stars represent $Sr_3As_4$, the triangles $Zn_3As_2$. (b) Crystal structure of $(Sr,Na)(Zn,Mn)_2As_2$. For $SrZn_2As_2$ $L_c$ =2.606Å, $L_{ab}$ = 2.549Å, α=111.834° and β=106.99°. (c) Lattice constants of $(Sr_{0.9}Na_{0.1})(Zn_{1-x}Mn_x)_2As_2$ for various Mn concentrations $x$.

**Fig.2.** (a) DC magnetization measured in $H$=500Oe in $(Sr_{1-y}Na_y)(Zn_{1-x}Mn_x)_2As_2$ with different charge doping levels $y$ and spin doping levels $x$. Inset shows the temperature dependence of the inverse susceptibility for $(Sr_{0.9}Na_{0.1})(Zn_{0.8}Mn_{0.2})_2As_2$. (b)Curie temperature $T_C$, Weiss temperature $θ$, effective paramagnetic moment $M_{eff}$ and saturation moment $M_{sat}$ for $(Sr_{0.9}Na_{0.1})(Zn_{1-x}Mn_x)_2As_2$. (c) Magnetic hysteresis curve M($H$) measured in $(Sr_{1-y}Na_y)(Zn_{0.85}Mn_{0.15})_2As_2$ at temperature $T$=2K for several different charge doping levels $y$. (d) Change of magnetization $M$ versus magnetic field $H$ measurements for the sample of $Sr_{0.8}Na_{0.2}(Zn_{0.85}Mn_{0.15})_2As_2$ for increasing temperature from 5K to 35K. (e) Arrott-Noakes plot under various temperature for the sample of $(Sr_{0.8}Na_{0.2})(Zn_{0.85}Mn_{0.15})_2As_2$.

**Fig.3**. (a) Temperature dependence of the resistivity $ρ$ at zero magnetic field of $(Sr_{1-y}Na_y)(Zn_{1-x}Mn_x)_2As_2$ for the pure Zn ($x$=0) and Mn 10% ($x$=0.1) systems with several different charge doping levels $y$. Note that the vertical axis for the pure $SrZn_2As_2$ is different from that for other specimens. (b) Magnetoresistivity $ρ(T)$ of $(Sr_{0.8}Na_{0.2})(Zn_{0.85}Mn_{0.15})_2As_2$ under various fields. (c) Negative magnetoresistance of $(Sr_{0.8}Na_{0.2})(Zn_{0.85}Mn_{0.15})_2As_2$ at different temperatures. (d) Hall resistivity of $(Sr_{0.8}Na_{0.2})(Zn_{0.85}Mn_{0.15})_2As_2$ at different temperatures, demonstrating p-type carriers with concentrations of $n \sim 10^{20}$ cm$^{-3}$ and the anomalous Hall effect due to spontaneous magnetization at $H = 0$. Inset shows the anomalous Hall effect at low field.

**Fig.4**. μSR measurements of $(Sr_{0.8}Na_{0.2})(Zn_{0.85}Mn_{0.15})_2As_2$: (a) Time spectra in zero field exhibiting the onset of rapid relaxation below $T=22K \sim 26$ K. The solid lines represent fits to the relaxation function (Eq.24 of Ref 28) for dilute spin systems in zero field for the static case. (b) Volume fraction of static magnetic order in ZF and spontaneous magnetization at $H=10$Oe. Inset shows temperature evolution of relaxation rate $a_s$, propotional to the ordered moment size. (c) Comparison between the static internal field parameter $a_s$ determined at $T = 2$ K by zero-field $\mu$SR versus the ferromagnetic Curie temperature $T_C$ observed in (Sr,Na)(Zn,Mn)$_2$As$_2$ (the present results with blue symbol), (Ga,Mn)As (Ref. 30), Li(Zn,Mn)As (Ref. 6), (Ba,K)(Zn,Mn)$_2$As$_2$ (Ref. 8), and (La,Ba)(Zn,Mn)AsO (Ref. 10). 4/3 is a factor multiplied to the parameter $a$ to adjust the difference from the simple exponential decay rate $\Lambda$ adopted in Ref. 30. The data point for the present work lies well above the common line shared by the earlier DMS systems, which suggests that the exchange interaction between Mn moments could be different than Ba122 and other DMS systems.

**Figure 1**

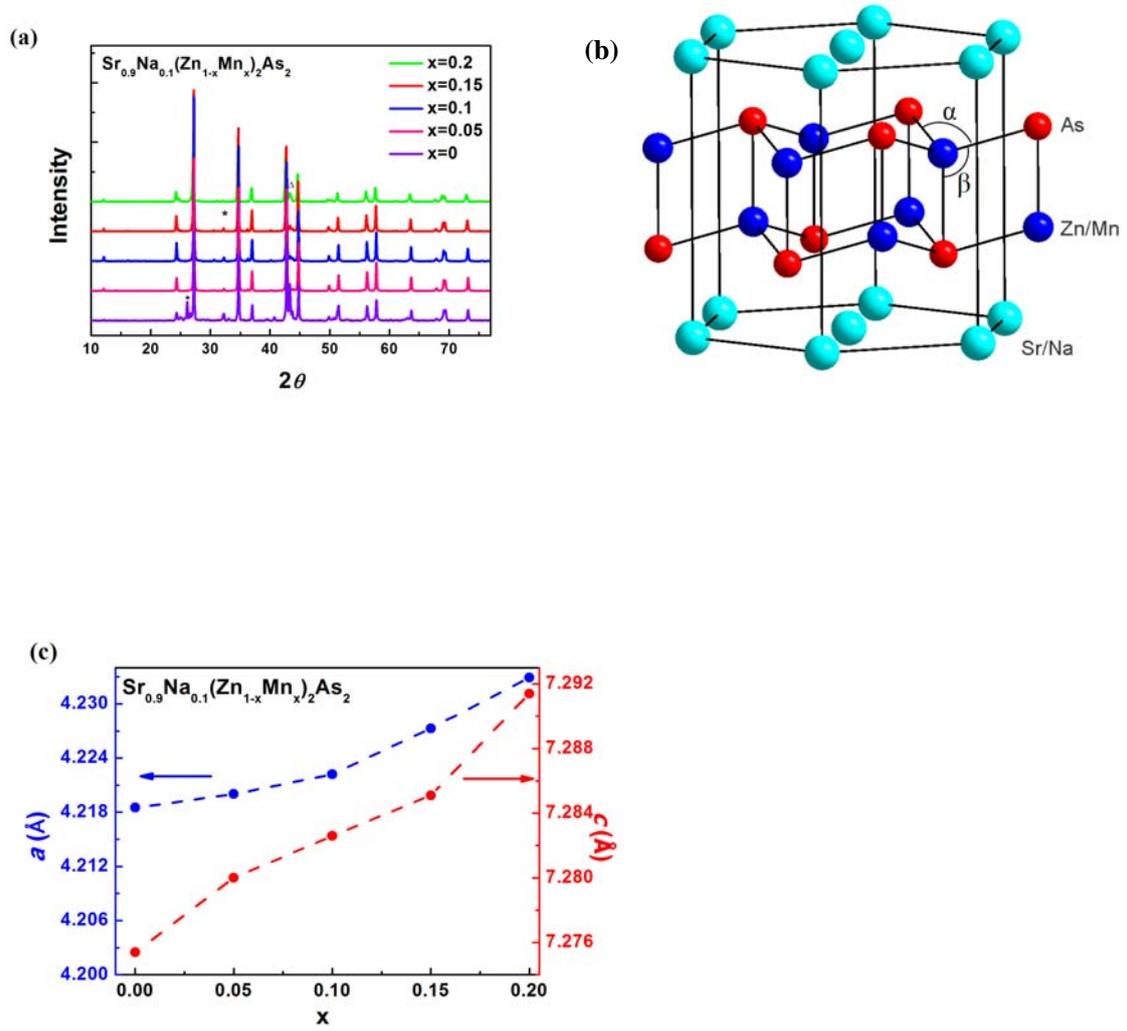

**Figure 2**

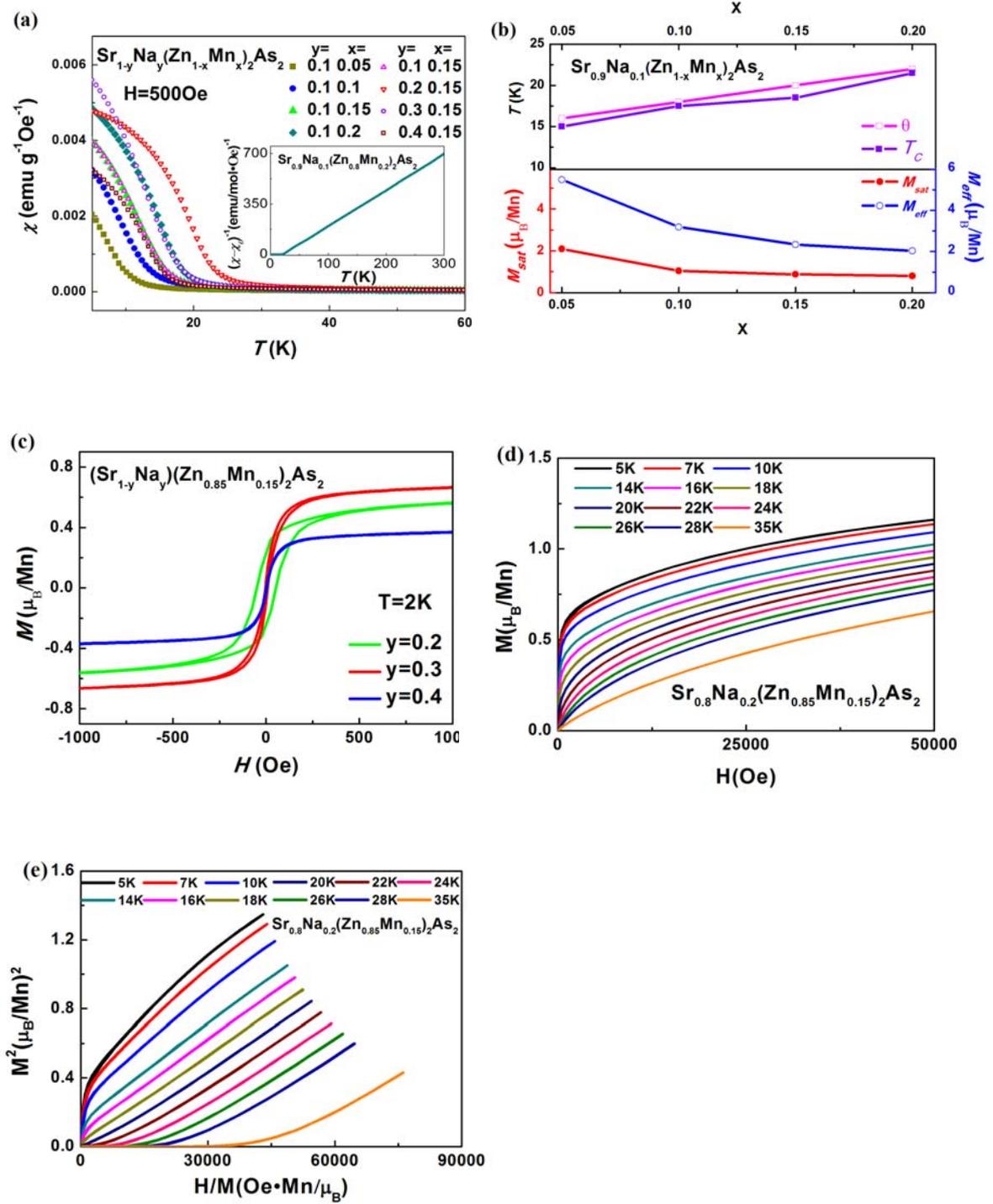

**Figure 3**

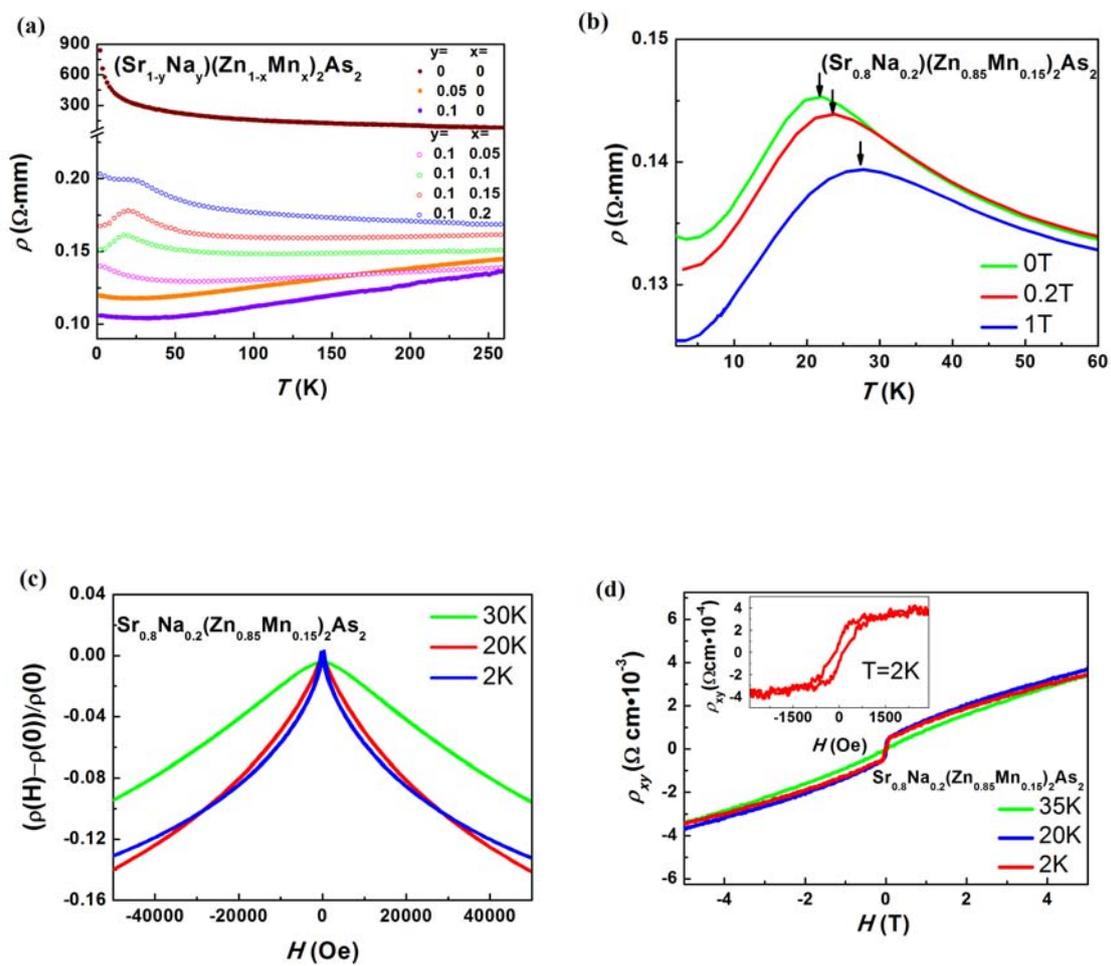

**Figure 4**

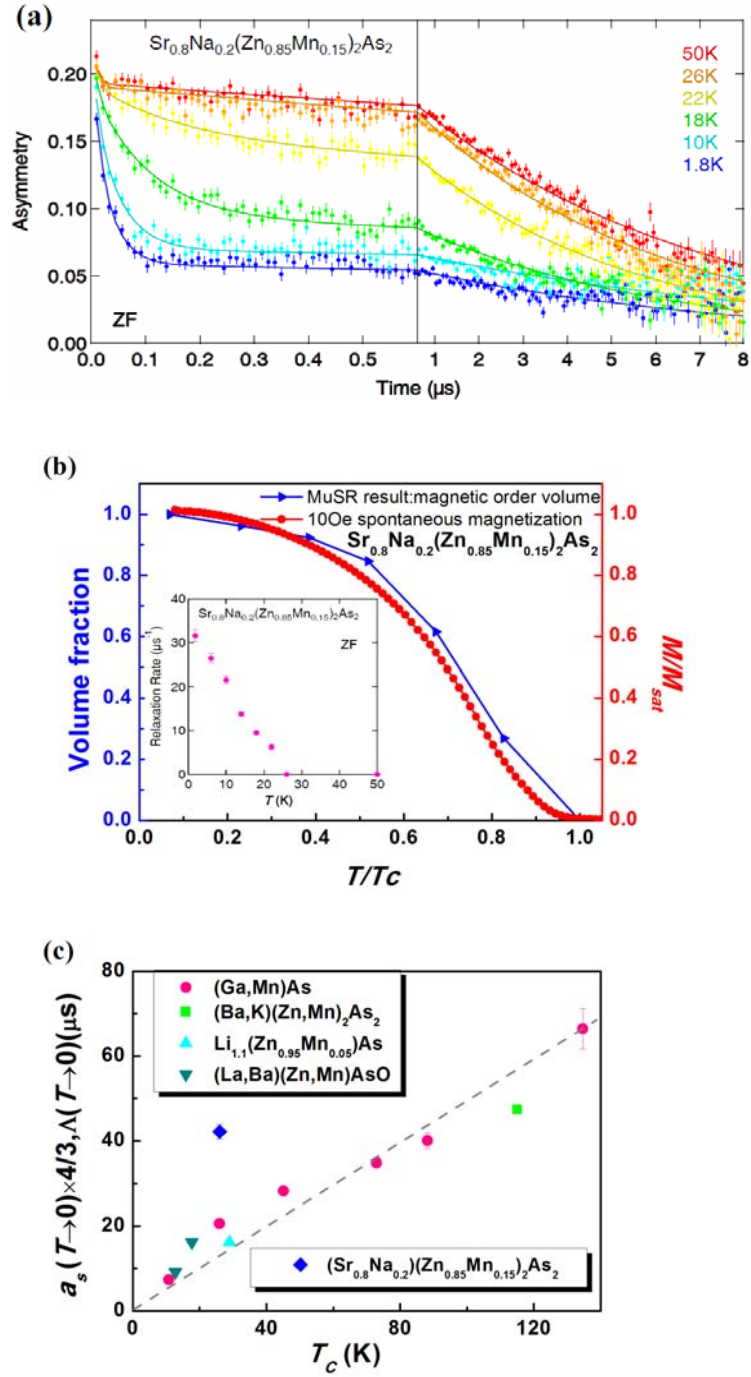